\documentclass[12pt]{article}

\usepackage{amsmath}
\usepackage{amssymb}
\usepackage{epsfig}

\makeatletter
\renewcommand\section{\@startsection {section}{1}{\z@}%
                                   {-3.5ex \@plus -1ex \@minus -.2ex}%
                                   {2.3ex \@plus.2ex}%
                                   {\normalfont\large\bfseries}}
\renewcommand\subsection{\@startsection{subsection}{2}{\z@}%
                                     {-3.25ex\@plus -1ex \@minus -.2ex}%
                                     {1.5ex \@plus .2ex}%
                                     {\normalfont\normalsize\bfseries}}
\renewcommand\subsubsection{\@startsection{subsubsection}{3}{\z@}%
                                     {-3.25ex\@plus -1ex \@minus -.2ex}%
                                     {1.5ex \@plus .2ex}%
                                     {\normalfont\normalsize\bfseries}}
\renewcommand\paragraph{\@startsection{paragraph}{4}{\z@}%
                                    {3.25ex \@plus1ex \@minus.2ex}%
                                    {-1em}%
                                    {\normalfont\normalsize\bfseries}}
\makeatother

\newcommand\as{\alpha_s}
\newcommand\asb{\bar\alpha_s}
\newcommand\calE{{\mathcal E}}
\newcommand\calG{{\mathcal G}}
\newcommand\calH{{\mathcal H}}
\newcommand\calR{{\mathcal R}}
\DeclareMathSymbol{\Pom}{\mathalpha}{AMSb}{"50}

\begin{document}

\begin{center}

{\LARGE\sf A Quantization of 2+1-Gravity\\
\vskip 0.2cm
Related to High-Energy Yang-Mills
Theory{\renewcommand{\thefootnote}{\fnsymbol{footnote}}\footnote{
Work supported in part by QCDNET contract FMRX-CT98-0194 and
by MIUR--Italy.
}}}\\
\end{center}
\vskip 0.1cm
\begin{center}
{\sc Marcello Ciafaloni$^{(a,b)}$}
\end{center}
\begin{center}
{\sc St\'ephane Munier$^{(a)}$}
\end{center}
\vskip 0.1cm
\begin{center}
{\em $^{(a)}$ Universit\`a di Firenze,
Dipartimento di Fisica\\
Via G.~Sansone 1, 50009 Sesto Fiorentino,
Italy}
\end{center}
\begin{center}
{\em $^{(b)}$ INFN, Sezione di Firenze, Italy}
\end{center}

\vskip 0.5cm

\begin{abstract}
We point out that canonical quantization of the two-body problem
in 2+1-Gravity is related to the high-energy equation in Yang-Mills
theory by a proper ordering of the relevant operators. This feature
arises from expanding the Hamiltonian around its conformal limit --
or treating running coupling effects in the Yang-Mills case -- and yields
a peculiar short distance behaviour of the wave functions.
\end{abstract}

\setcounter{footnote}{0}

\section{Introduction}

It is a common feature of several theoretical models
to show a (nearly) two-dimensional dynamics
either because of fundamental reasons (as in string theory)
or due to the importance of certain kinematical configurations.
Here we focus on two models of the latter class, namely gauge theories
in the high-energy limit and 2+1 gravity with pointlike matter.
It is known that both models possess a 2+1 dimensional ``configuration''
space, with a ``time'' parameter of quite different meaning in the two
cases, and two space dimensions (the transverse ones to the 
high energy momenta
in the gauge case). It is less known, and we emphasize it here that
the Hamiltonian of either model is eventually provided by the same
operator and is related to a fundamental scale of the system. 

In other words, in this paper
we show that Hamiltonian quantization relates in a 
nontrivial manner a well-known high-energy tool -- the so-called
BFKL equation \cite{Kuraev:1977fs,Balitsky:1978ic} -- 
to a quantized (two-body) gravitational system
with proper ordering of the relevant operators. Perhaps the reason why
two such different systems happen to have eventually the same
dynamics is that both admit a conformal limit in which the Hamiltonian
is just the dilatation operator. Deviations from the conformal behaviour
are then treated by expanding in some mass parameter (with different 
physical meaning in the two cases).

It should be noticed that the relationship of gravitational theories
to gauge theories and/or conformal theories may have a more
general origin \cite{Maldacena:1998re,Strominger:2001pn}. For
gravity in 2+1 dimensions \cite{Deser:1984tn}
this appears from the relevance of Liouville fields in the
classical solutions with cosmological constant 
\cite{Deser:1984dr,Klemm:2002ir}
and/or with matter sources \cite{Bellini:1995rw,Menotti:1999pn}.
It is not known however whether these suggestions may have
an impact at quantum level also.

From the point of view of 2+1-Gravity, the quantization of
the two-body system proposed here provides an alternative to
the standard one of Deser, Jackiw and 't~Hooft 
(DJH)~\cite{Deser:1984tn,Deser:1988qn,'tHooft:1988yr},
which adds up to those already known \cite{Carlip:1993zi}.
The large-distance properties (in particular phase-shifts and scattering
angle) stay the same, but the wave function behaves differently
at short distances where it shows an anomalous dimension
behaviour as in the gauge theory case.

After introducing Hamiltonian and physics of our two models
in Secs.~2 and~3, we discuss in detail the ensuing quantum
properties in Sec.~4 and possible developments at many body level
in the conclusive section~5.

\section{The 2+1-Gravity Hamiltonian}

Gravity in three space-time dimensions is characterized by the fact
that the Riemann tensor is proportional to the Einstein tensor and thus
to the energy-momentum tensor. As a consequence, space-time is flat
outside the matter sources. If the latter are pointlike particles,
local Minkowskian coordinates can be extended all around them, but
are in general multivalued, i.e. carry nontrivial monodromy
transformations for parallel transport in a closed loop around
each particle site \cite{Witten:1988hc,Achucarro:1986vz}.

For a spinless particle at rest, the loop integral of the connection
is $8\pi Gm$ -- or just $m$, in units of the energy $1/8\pi G$,
so that the Minkowskian coordinates $X^a=(T,Z,\bar{Z})$ possess
a branch-cut characterized by the discontinuity relation
\begin{equation}
Z_{II}=e^{-im}Z_{I},\ T_{II}=T_{I}\ \ (Z=X+iY)
\label{eq:monotriv}
\end{equation}
between values above and below the cut. This corresponds to a cut-out
sector or deficit angle $m$
\begin{equation}
|\Theta|=|\mbox{arg}\;Z|<\pi\alpha,\ \alpha=1-\frac{m}{2\pi}=1-4Gm
\end{equation}
typical of a conical space. For moving particles of momenta $P_n^a$,
the relation (\ref{eq:monotriv}) is boosted to Lorentz transformations
-- the DJH matching conditions \cite{Deser:1984tn}
\begin{equation}
(X_{II}-X_n)^a=L(P_n)_b^a(X_I-X_n)^b,\ (P_n^2=m_n^2,\ n=1,...,N)
\label{eq:monoboost}
\end{equation}
and the latter {\em do not commute} for nonvanishing relative
velocities.

In the static case (\ref{eq:monotriv}), it is straightforward
to construct single-valued coordinates $x^\mu=(t,z,\bar{z})$
around each particle, by a coordinate transformation of the type
\begin{equation}
T=t,\ Z=z\left(\frac{z}{\lambda}\right)^{-\mu}\sim z^\alpha,\ 
(\mu=\frac{m}{2\pi}=4Gm)\ ,
\end{equation}
so that $Z\rightarrow e^{-im}Z$ when $z\rightarrow e^{2i\pi}z$.
The corresponding metric is nontrivial, with line element
\begin{equation}
ds^2=dt^2-\alpha^2\left|\frac{z}{\lambda}\right|^{-2\mu}|dz|^2,
\ (|\mbox{Arg}\ z|\leq\pi)
\label{eq:metric}
\end{equation}
and yields the conformal-gauge description.
Finally, the scale change $\rho=\lambda(r/\lambda)^\alpha$
(we denote by $r$ and $\theta$ the modulus and argument of $z$)
brings Eq.~(\ref{eq:metric}) into the canonical DJH form
\begin{equation}
ds^2=dt^2-d\rho^2-\alpha^2\rho^2 d\theta^2
\label{eq:metricDJH}
\end{equation}
in which the geometry of a cone with aperture
$2\pi\alpha=2\pi-2\pi\mu$ is transparent.

We have introduced in the above equations a scale parameter
$\lambda$ which is arbitrary at this stage, but becomes
dependent on the dynamical variables of the system
if the metric (\ref{eq:metric}) is interpreted as the
asymptotic metric\footnote{
Actually, the asymptotic metric takes up a more general
ADM form \cite{ADM} if the angular momentum of the system
is nonvanishing \cite{Deser:1984tn}.
In Ref.~\cite{Moncrief:1989dx}, the case of a more complicated topology
of space was also investigated.} for $|z|\gg\lambda(z_n,p_n)$ of a
system of many particles of coordinates $\{z_n,p_n\}$
and invariant mass $M=2\pi\mu$.

The simplest quantization procedure is that of a test
particle in the conical space of Eqs.~(\ref{eq:metric})
and (\ref{eq:metricDJH}), described by the eigenfunctions
of the Laplace operator on the cone~\cite{Deser:1988qn,'tHooft:1988yr}
\begin{equation}
\Delta_{|\mbox{\tiny cone}}=\left|\frac{z}{\lambda}\right|^{2\mu}
\Delta_z=\left|\frac{z}{\lambda}\right|^{2\mu}|p|^2\ ,
\label{eq:laplace}
\end{equation}
where $p=-i\partial_z$. Generalizing such approach to the dynamical
many-body system \cite{Bellini:1996mb} is nontrivial, 
due to the noncommutativity
of the monodromies (\ref{eq:monoboost}).
At the classical level, a single-valued metric of conformal
type was obtained in the instantaneous gauge 
\cite{Bellini:1995rw}
by Bellini, Valtancoli and one of us (M.C.) \cite{Bellini:1996mb}, 
and by
Welling \cite{Welling:1996er}. The same solution was exhibited by Menotti
and Seminara \cite{Menotti:1999pn,Cantini:2000dd} 
in a canonical formalism, which allows
the derivation of the classical two-body Hamiltonian\footnote{
In a general time gauge the Hamiltonian may differ from 
(\ref{eq:hamiltoniangravity}) by a constant factor,
which for a ``local'' observer is 
$\sim (\mu\!-\!\mu_1\!-\!\mu_2)$, as in the equations of motion
of Ref.~\cite{Bellini:1996mb}.}
\begin{equation}
\begin{split}
\calH&=\mu\log|z|^2+\log|p|^2\ \ (\mu=\frac{M}{2\pi})\\
     &=\mu\log\lambda^2(z,p)\ .
\end{split}
\label{eq:hamiltoniangravity}
\end{equation}
Here $z=z_2-z_1$ is the relative coordinate of the two particles,
and $p$ is the conjugate momentum, while $\lambda(z,p)$
turns out to be \cite{Cantini:2000dd} the scale parameter introduced
in Eq.~(\ref{eq:metric}), interpreted as the
asymptotic metric of the 2-body system.

The expression (\ref{eq:hamiltoniangravity}) is formally
the logarithm of (\ref{eq:laplace}) and thus may lead to a
quantization much similar to that of Deser, Jackiw and 't Hooft.
However, the decomposition (\ref{eq:hamiltoniangravity}) as a sum
of two contributions suggests a quantization with different ordering,
related to the BFKL equation with running coupling, as we shall see
in the following.

\section{The High-Energy Evolution Equation}

The basic question of high-energy QCD is to find the total
cross section $\calG(k,k_0;Y)$ for the scattering of
two gluons at scales $k,k_0$ and relative rapidity $Y=\log(s/(kk_0))$,
$s$ being their center-of-mass squared energy. The pioneering
work of Balisky, Fadin, Kuraev and Lipatov \cite{Kuraev:1977fs,
Balitsky:1978ic} showed that
the perturbative high-energy behaviour could be resummed in the
leading logarithmic approximation by an evolution equation
involving a two dimensional Hamiltonian
$\calH_0$ \cite{Lipatov:1990zb,Lipatov:1993qn}
\begin{equation}
\begin{split}
-\frac{\partial}{\partial Y}\calG(Y)&=\asb\calH_0\calG(Y)\\
\calG(k,k_0;0)&=\delta^2(k-k_0)\ ,
\end{split}
\label{eq:bfkl}
\end{equation}
where
\begin{equation}
\calH_0=\log|k|^2+\log|z|^2-2\psi(1)\ ,
\label{eq:hamiltonianbfkl0}
\end{equation}
$k=k_1+ik_2$, $z\equiv i\partial_k=i(\partial_1-i\partial_2)/2$ 
is the variable conjugated to the 2-dimensional 
momentum $k$  (its Hermitian conjugate operator is
$\bar z\equiv i\partial_{\bar k}$)
and $\asb=\as N_c/\pi$ is the QCD squared coupling constant.

The BFKL evolution equation (\ref{eq:bfkl}) is solved in terms of the
eigenfunctions of the Hamiltonian (\ref{eq:hamiltonianbfkl0}),which is
actually scale-invariant, so that it has
characteristic function $-\chi_n(\nu)$,
where
\begin{equation}
\chi_n(\nu)=2\psi(1)-2\calR e\,\psi\left(\frac{1\!+\!|n|}{2}
\!+\!i\nu\right)=2\psi(1)-2\calR e\,\psi\left(\frac{1\!-\!|n|}{2}
\!+\!i\nu\right)
\label{eq:characteristic}
\end{equation}
on the power-behaved eigenfunctions 
$\Psi_{n,\nu}(k,\bar k)\!=\!
k^{\gamma-1}\bar{k}^{\tilde\gamma-1}$ ($\gamma\!=\!
\tilde\gamma\!-\!n\!=\!(1\!-\!n)/2\!+\!i\nu$
and $\psi(x)=\Gamma^\prime(x)/\Gamma(x)$).
Thus, the evolution equation~(\ref{eq:bfkl}) and 
Eq.~(\ref{eq:characteristic}) show that the 2-gluon correlator is
exponentially behaved in $Y$ (or power-behaved in $s$), with
exponent $\omega_\Pom=\asb\chi_0(0)$, yielding the so-called
hard Pomeron behaviour of the cross section.

Let us outline the proof of these classical results.
The fact that the eigenfunctions $\Psi_{n,\nu}$
are powers of $k$ and $\bar k$ is due to scale invariance of
the Hamiltonian~(\ref{eq:hamiltonianbfkl0}).
Single-valuedness of the cross section imposes that
$\tilde\gamma\!-\!\gamma$ be an integer. 
The eigenvalues (\ref{eq:characteristic})
are then computed by using the relation of $\calH_0$ to the
sum of $\log|k|^2$ and $\log|\partial_k|^2$ which defines
our ordering.
We start from the definition of the $\log|z|^2=\log|\partial_k|^2$ 
operator acting on a given function $\Psi(k,\bar k)$:
\begin{equation}
\log|z|^2 \Psi(k,\bar k)=\int\frac{d^2\rho}{(2\pi)^2}
e^{i(k\bar\rho+\bar k\rho)/2}
\log\left(\frac{|\rho|^2}{4}\right)\tilde \Psi(\rho,\bar\rho)\ ,
\label{eq:technique1}
\end{equation}
where $\tilde \Psi(\rho,\bar\rho)$ is the Fourier-transform of 
$\Psi(k,\bar k)$.
It is useful to take the following representation for the 
logarithm:
\begin{equation}
\log\left(\frac{|\rho|^2}{4}\right)=-\frac{d}{d\epsilon}_{|\epsilon=0}
\left(\frac{|\rho|^2}{4}\right)^{-\epsilon}\ .
\end{equation}
We specialize to the power-functions
$\Psi_{n,\nu}(k,\bar k)=k^{\gamma-1}\bar k^{\tilde\gamma-1}$. 
Their Fourier transform 
$\tilde \Psi_{n,\nu}(\rho,\bar\rho)$ 
can be computed using the formula
\begin{equation}
\int d^2 k\,k^{\gamma-1}\bar k^{\tilde\gamma-1}e^{-ik\cdot\rho}
=\pi\, i^{\gamma-\tilde\gamma}
\left(\frac{2}{\rho}\right)^{\tilde\gamma}
\left(\frac{2}{\bar\rho}\right)^{\gamma}
\frac{\Gamma(\tilde\gamma)}{\Gamma(1\!-\!\gamma)}\ ,
\end{equation}
which holds for $\tilde\gamma\!-\!\gamma$ positive (the case 
$\tilde\gamma\!-\!\gamma$ negative
is obtained by exchanging $\gamma$ and $\tilde\gamma$).
The integral over $\rho$ in Eq.~(\ref{eq:technique1}) 
is performed using again
the same formula. Finally one takes
the derivative with respect to $\epsilon$ to obtain
\begin{equation}
(\log|z|^2+\log|k|^2)\Psi_{n,\nu}(k,\bar k)
=(\psi(\gamma)+\psi(1-\tilde\gamma))
\Psi_{n,\nu}(k,\bar k)\ .
\label{eq:rellogpsi}
\end{equation}
As $1-\tilde\gamma=\bar\gamma$, this equation proves that
the functions 
$\Psi_{n,\nu}(k,\bar k)\!=\!
k^{\gamma-1}\bar k^{\tilde\gamma-1}$ are eigenfunctions
of $\calH_0$ with eigenvalues given by Eq.~(\ref{eq:characteristic}).
Note that equation~(\ref{eq:rellogpsi}) 
can be formally written as an identity between operators:
$\log|\partial_k|^2\!+\!\log|k|^2\!=\!\psi(1\!+\!k\partial_k)
\!+\!\psi(-\bar k\partial_{\bar k})$. 
Consequently, 
the BFKL Hamiltonian can be expressed as a function of the
dilatation operators $k\partial_k$ and $\bar k\partial_{\bar k}$,
besides admitting the customary integral operator
formulation \cite{Kuraev:1977fs,Balitsky:1978ic}.

Finally, solutions to the BFKL evolution~(\ref{eq:bfkl})
are linear combinations of the eigenfunctions
to which one applies the evolution operator
\begin{equation}
\begin{split}
\calG(k,k_0;Y)&=\sum_n\int\frac{d\nu}{2\pi}\,
\bar\Psi_{n,\nu}(k,\bar k)\,e^{-\asb\calH_0 Y}
\Psi_{n,\nu}(k_0,\bar k_0)\\
&=\frac{1}{|k k_0|}
\sum_n
\int\frac{d\nu}{2\pi}
\,
\left|\frac{k}{k_0}\right|^{-2i\nu}
\left(\frac{k}{\bar k}
\frac{\bar k_0}{k_0}\right)^{n/2}
e^{\asb\chi_n(\nu) Y}\ .
\end{split}
\end{equation}
The dominant large $Y$ behaviour is given by the saddle point
of the azimuthally symmetric $n\!=\!0$ component, which lies
at $\nu\!=\!0$ so that 
$\calG(Y){\sim}\exp({\asb\chi_0(0)\cdot Y})$, as stated before.


The model just outlined shows conformal invariance at
non-vanishing momentum transfer, and can be generalized
to many-gluon correlators \cite{Lipatov:1993qn} and connected to
an integrable 2-dimensional model \cite{Faddeev:1995zg}.\\

When subleading logs corrections \cite{Fadin:1998py,Ciafaloni:1998gs,
Camici:1997ij}
are taken into account,
the picture changes qualitatively, because at this level the
QCD coupling acquires renormalization group (RG) evolution
in the form
\begin{equation}
\asb(k^2)=\frac{1}{b\log k^2/\Lambda^2}\ ,
\label{eq:running}
\end{equation}
where $\Lambda^2$ is the RG invariant QCD scale.
Therefore, by introducing the variable $\omega$ conjugated
to $Y$, Eq.~(\ref{eq:bfkl}) takes the form
\begin{equation}
\left[(1+b\omega)\log|k|^2+\log|z|^2\right]\calG=
b\omega\log\Lambda^2 \calG\ ,
\label{eq:bfklrunning}
\end{equation}
apart from a delta function source $b\log(k_0^2/\Lambda^2)
\delta^2(k-k_0)$.
Notice that Eq.~(\ref{eq:bfklrunning}) is no longer scale-invariant,
and involves a new kind of Hamiltonian, of the type
\begin{equation}
\calH=\log|k|^2+\mu\log|z|^2\ \ \ \ \ \ \ (\mu=\frac{1}{1+b\omega})
\label{eq:hamiltonianbfkl}
\end{equation}
whose eigenvalues are related to $b\omega$ and to the QCD scale
$\Lambda^2$ itself.
This Hamiltonian is exactly 
the same as the one given in Eq.~(\ref{eq:hamiltoniangravity}),
except that (\ref{eq:hamiltonianbfkl}) is already quantized: indeed
$k$ and $z$ are operators satisfying to the commutation relation
$[z,k]=i$.

Let us solve the eigenvalue equation $\calH\phi_E=E\phi_E$
for a coordinate-dependent wave-function $\phi_E(z,\bar z)$.
Although the Hamiltonian is no more scale invariant,
we shall take the functions $\Psi_{n,\nu}(z,\bar z)
=z^{-\gamma}
\bar z^{-\tilde\gamma}/(\pi\sqrt{2})$ 
as a basis for its eigenfunctions,
where $\gamma(n,\nu)$ and 
$\tilde\gamma(n,\nu)$
were introduced in the previous section,
thus spanning an $L_2$ space. 
The normalization is chosen
so that the $\Psi_{n,\nu}$ are orthonormal
$\int d^2 z\,\Psi_{n,\nu}(z,\bar z)
\bar\Psi_{n^\prime,\nu^\prime}(z,\bar z)\!=\!\delta_{n,n^\prime}
\delta(\nu\!-\!\nu^\prime)$.
Since the $\Psi$'s are eigenfunctions of $\calH(\mu\!=\!1)$
with the eigenvalues in Eq.~(\ref{eq:rellogpsi}), we can set
$\calH=\calH(\mu\!=\!1)\!+\!(\mu\!-\!1)\log|z|^2$
and we notice the representation $\log|z|^2\rightarrow
-\partial/\partial(i\nu)$.
It is then easy to see that the linear combination
\begin{equation}
\phi_E(z,\bar z)=\sum_n\int\frac{d\nu}{2\pi}\Psi_{n,\nu}(z,\bar z)
f_{E\,n,\nu}
\label{eq:eigen}
\end{equation}
is an eigenfunction of $\calH$ if the coefficients of
the expansion are:
\begin{equation}
f_{E\,n,\nu}=\sqrt{\frac{2\pi}{1\!-\!\mu}}
\exp\left(-\frac{i\nu E}{1-\mu}-\frac{X_n(\nu)}{1-\mu}\right)\ ,
\end{equation}
where
\begin{equation}
X_n(\nu)=\int_0^{i\nu} d({i\nu^\prime})\left(\chi_n(\nu^\prime)
-2\psi(1)\right)
=\log\frac{\Gamma\left(\frac{1\!+\!|n|}{2}\!-\!i\nu\right)}
{\Gamma\left(\frac{1\!+\!|n|}{2}\!+\!i\nu\right)}\ .
\end{equation}
The normalization has been fixed by
requiring that the set of functions 
$\phi_E$ be orthonormal:
$\int d^2z \phi_E(z,\bar z)\bar\phi_{E^\prime}(z,\bar z)
=\delta(E\!-\!E^\prime)$.

The method just outlined 
is the so-called ``$\gamma$-representation'' widely used in
high-energy physics. The expansion (\ref{eq:eigen}) is
the natural equivalent of a Fourier expansion in the case in which the
Hamiltonian is a function of the dilatation operator $z\partial_z$.
The set of functions $\Psi_{n,\nu}(z,\bar z)$ for the expansion
plays the role of the Fourier basis $e^{i k\cdot z}$:
the latter are eigenfunctions of the translation operator
$\partial_z$ while the former are
eigenfunctions of the dilatation operator $z\partial_z$.
Finally, the energy $E$ is fixed in this case
by Eq.~(\ref{eq:bfklrunning})
to be $E\!=\!(1\!-\!\mu)\log\Lambda^2$ and is
thus related to the QCD scale.

\section{Quantum Scattering Solutions}

We have just shown that the cross section for
gluon scattering at high energy obeys a Schr\"odinger equation.
Its Hamiltonian is classically the same as the one describing
the diffusion of 2 massive particles in 2+1 dimensions (see section~2).

We shall now
take advantage of this equivalence
to investigate the new quantization scheme for gravity
which comes from the ordering induced by Eq.~(\ref{eq:rellogpsi}).
We shall study the properties of the wave function
obtained in the previous section.

We will consider separately each component of given angular
momentum $n$ of the wave function.
Introducing
the modulus $r$ and argument $\theta$ of the transverse
coordinate vector $z$, we write
\begin{equation}
\phi_E(r,\theta)=\sum_n\varphi_{n}(r,\theta|E)\ .
\end{equation}
We recast the
partial waves in the following form:
\begin{equation}
\varphi_n(r,\theta)=\frac{e^{in\theta}}{\sqrt{\pi(1\!-\!\mu)}}
\frac{1}{r}\int_{-\infty}^{+\infty}\frac{d\nu}{2\pi}e^{\calE_n(i\nu|r)}\ ,
\label{eq:wave0}
\end{equation}
where
\begin{equation}
\calE_n(i\nu|r)=-2i\nu\log(\kappa(E) r)\!-\!
\frac{1}{1\!-\!\mu}\log\frac{\Gamma\left(
\frac{1\!+\!|n|}{2}\!-\!i\nu\right)}
{\Gamma\left(\frac{1\!+\!n}{2}\!-\!i\nu\right)}
\end{equation} 
and where we have singled out the scale of distances by defining
\begin{equation}
\kappa^2(E)=e^{E/(1\!-\!\mu)}\ .
\end{equation}
For sake of simplicity, the 
$E$-dependence will be implicit in most of
the following equations which involve $\kappa$.

We note that in the particular case $\mu=0$, the
wave function~(\ref{eq:wave0}) reduces to a Bessel function:
\begin{equation}
\varphi_{n}(r,\theta)|_{\mu=0}=
\frac{\kappa e^{i n\theta}}{\sqrt{\pi}}
J_{|n|}(2\kappa r)\ .
\end{equation}
For the general case,
we will investigate the behaviour of the wave function~(\ref{eq:wave0}) 
in different limits of 
the parameter $\kappa r$.
Technically, we will use the steepest descent method.
The equation 
$\partial\calE_n(i\nu|r)/\partial(i\nu)\!=\!0$
defines the saddle points
$i\nu_s$ as roots of the equation
\begin{equation}
-2\log(\kappa r)\!+\!\frac{1}{1\!-\!\mu}\left\{\psi
\left(\frac{1\!+\!|n|}{2}\!+\!i\nu_s\right)
\!+\!\psi\left(\frac{1\!+\!|n|}{2}\!-\!i\nu_s\right)\right\}=0\ .
\label{eq:sadeq}
\end{equation}
As we shall only be interested
in the leading terms in $|\log(\kappa r)|$,
we will solve this equation by taking the relevant approximations
for the polylogarithm function $\psi$.
The wave function reads in the saddle point approximation
\begin{multline}
\varphi_n(r,\theta)\simeq\frac{e^{in\theta}}{\sqrt{\pi(1\!-\!\mu)}}
\frac{1}{r}\cdot
\sum_{i\nu_s}
\frac{e^{\calE_n(i\nu_s|r)}}{\sqrt{2\pi
(\partial^2\calE_n(i\nu|r)/\partial (i\nu)^2)}|_{i\nu=i\nu_s}}\\
\times\left\{1+3
\frac{\partial^4\calE_n(i\nu|r)/\partial (i\nu)^4|_{i\nu=i\nu_s}}
{(\partial^2\calE_n(i\nu|r)/\partial(i\nu)^2)^2|_{i\nu=i\nu_s}}
+...\right\}\ .
\label{eq:saddle}
\end{multline}
The sum goes over all the roots of Eq.~(\ref{eq:sadeq}).
We have singled out the dominant contribution as well as 
the first correction to it.\\

First, the study of the large distance behaviour of the wave function 
($\kappa r\!\gg\! 1$) enables
to identify the phase shift due to the scattering \cite{Deser:1988qn}.
Since $\log(\kappa r)$ is large and positive, the saddle point 
defined by Eq.~(\ref{eq:sadeq})
sits at large $i\nu_s$. One can use the approximation
$\psi(z){\sim} \log z$ which then gives 
$(\kappa r)^{2(1-\mu)}\!=\!((1\!+\!|n|)/2)^2\!+\!\nu_s^2$
up to terms of relative order $1/(\kappa r)^{(1-\mu)}$.
Taking into account the two roots of this equation,
we use formula (\ref{eq:saddle})
and we obtain
\begin{equation}
\varphi_n(r,\theta)\underset{\kappa r\gg 1}{\simeq}
\frac{\kappa e^{in\theta}}{\pi}
(\kappa r)^{-(1+\mu)/2}
\cos\left(\frac{2(\kappa r)^{1-\mu}}
{1\!-\!\mu}-\frac{|n|\pi}{2(1\!-\!\mu)}
-\frac{\pi}{4}
\right)
\label{eq:scat}
\end{equation}
for the leading contribution.
The first relative correction
is $3/4\cdot(1\!-\!\mu)(\kappa r)^{-(1\!-\!\mu)}$ 
from Eq.~(\ref{eq:saddle}). It is subleading
in $|\log(\kappa r)|$, which justifies the method.

This result should be compared to the wave function
for scattering in flat space:
the partial wave of angular momentum $n$
is given by
$J_{|n|}(2\kappa r)\sim \sqrt{1/\pi\kappa r}
\cos(2\kappa r\!-\!|n|\pi/2\!-\!\pi/4)$.
In this case, the wave front is rotated by an angle $\pi$
in the scattering process.
In our case, by analogy, we see 
on Eq.~(\ref{eq:scat})
that the wave front
is rotated by $\pi/(1\!-\!\mu)$.
Thus the scattering angle is $\pi\mu/(1\!-\!\mu)$,
and corresponds to the deficit angle of the effective conical
space in which the particles are moving.
Both phase shifts in Eq.~(\ref{eq:scat}) and scattering
angle agree with the results of DJH.\\

The small distance behaviour of the wave function is also
of interest because
in the context of
high energy scattering,
this regime corresponds to a configuration in which
the interacting gluons have large virtualities.
As $\log(\kappa r)$ is large and negative, the saddle point 
defined by Eq.~(\ref{eq:sadeq})
sits near the pole of the $\psi$ function at
$(1\!+\!|n|)/2\!+\!i\nu\!=\!0$. The polylogarithm 
functions are approximated
by $\psi(z)\!\sim\! -1/z$.
The equation for the saddle point then
gives $(1\!+\!|n|)/2\!+\!i\nu_s\!=\!-1/(2(1\!-\!\mu)\log(\kappa r))$.
Applying once again formula~(\ref{eq:saddle}), one obtains
the leading term
\begin{equation}
\varphi_n(r,\theta)\underset{\kappa r\ll 1}{\simeq}
\frac{\kappa e^{in\theta}}{2\sqrt{2}\pi(1\!-\!\mu)}
\left(\frac{2e(1\!-\!\mu)}
{\Gamma(1\!+\!|n|)}\right)^{1/(1\!-\!\mu)}
(\kappa r)^{|n|}
|\log(\kappa r)|^{\mu/(1\!-\!\mu)}\ .
\label{eq:smalldis}
\end{equation}
A bit of care is in order in this case.
Indeed, the first relative correction to this
approximation (second term in the parenthesis in
Eq.~(\ref{eq:saddle}))
is $18(1\!-\!\mu)$.
Hence the saddle point method only gives
the leading behaviour of the wave function
up to terms of relative order
$1\!-\!\mu$.

We note that in the particular case
in which $1/(1\!-\!\mu)$ is an integer,
the wave function is a Meijer function.
All subleading orders can be computed
by expanding it in a power series of $\kappa r$.
This can be done by picking up 
the successive poles of the integrand in the upper $\nu$-plane.
For $\mu=1/2$, we obtain the following result:
\begin{multline}
\varphi_n(r,\theta)=
{2\kappa}e^{in\theta}\sqrt{\frac{2}{\pi}}(\kappa r)^{|n|}
\left(\frac{|\log(\kappa r)|}{\Gamma^2(1\!+\!|n|)}
{}_0F_3(1,1\!+\!|n|,1\!+\!n,(\kappa r)^2)\right.\\
\left.+\sum_{k=0}^\infty
(\kappa r)^{2k}\frac{\psi(1\!+\!k)\!+\!\psi(1\!+\!|n|\!+\!k)}
{\Gamma^2(1\!+\!k)\Gamma^2(1\!+\!|n|\!+\!k)}\right)\ .
\label{eq:partcase}
\end{multline}

For $\mu\ne 0$, the behaviour of the wave function
for small $\kappa r$ computed in 
Eqs.~(\ref{eq:smalldis},\ref{eq:partcase}) is
to be contrasted with the one
found in Refs.~\cite{Deser:1988qn,Cantini:2000dd},
within the DJH quantization scheme
\begin{equation}
\varphi_n(r,\theta)=\frac{e^{in\theta}}{\sqrt{2\pi}}
J_{\frac{|n|}{1\!-\!\mu}}\left(
\frac{\left(\kappa r\right)^{1\!-\!\mu}}{1\!-\!\mu}
\right)
\underset{\kappa r\ll 1}{\simeq} 
\frac{e^{in\theta}}{\sqrt{2\pi}}
\frac{(\kappa r)^{|n|}}
{2^{\frac{|n|}{1\!-\!\mu}}\Gamma
\left(1\!+\!\frac{|n|}{1\!-\!\mu}\right)}
\end{equation}
which does not show the logarithmic corrections of 
Eqs.~(\ref{eq:smalldis},\ref{eq:partcase}).
The latter are due to the Yang-Mills anomalous dimension
which in turn are embodied in the small $\gamma$ behaviour of the
eigenvalue function~(\ref{eq:rellogpsi}).

It is amusing to note that the relevant regimes of the
two theories are somehow exchanged. The short distance
behaviour, which in 2+1-Gravity is dependent on the quantization
procedure and is expected to be of strong-coupling nature is
related to the perturbative anomalous dimension regime of the
gauge model. On the other hand, the large distance behaviour,
fixed in 2+1-Gravity by the semiclassical limit, is related to the
large $|\mbox{Im}\,\gamma|$ behaviour of the characteristic
function, which in the gauge model is expected to be affected
by higher order corrections. This interchange of weak- and
strong-coupling regimes is analogous to what happens in
duality transformations.\\

Finally, we check the consistency of the wave function
that we have constructed by taking its classical limit.
We get the time evolution of the system by
applying the operator $e^{i\calH t}$ to a wave packet.
We choose to construct the latter by combining 
all the wave functions corresponding
to different values of the energy $E$ with an equal weight:
$\Phi_n(t,r,\theta)=e^{i\calH t}\int_{-\infty}^{+\infty} 
dE \phi_n(r,\theta|E)$.
This enables to express the variable $\nu$ as a function of 
$t$:
$\nu=(1\!-\!\mu)t$ and therefore to get rid of the integration
over $\nu$.
We obtain: 
\begin{multline}
\Phi_n(t,r,\theta)=e^{in\theta}
\sqrt{\frac{1\!-\!\mu}{\pi}}
\exp\Bigg\lbrace
-(1\!+\!2it(1\!-\!\mu))\log r\!\\
-\!\frac{1}{1\!-\!\mu}
\log\frac{\Gamma(\frac{1\!+\!|n|}{2}\!+\!i(1\!-\!\mu)t)}
{\Gamma(\frac{1\!+\!|n|}{2}\!-\!i(1\!-\!\mu)t)}
\Bigg\rbrace\ .
\end{multline}
The classical limit of the wave function
corresponds to large angular momenta, i.e. large $|n|$.
In this limit, the phase is stationary for:
\begin{equation}
\begin{split}
(1\!-\!\mu)\theta&=\tan^{-1}\frac{2(1\!-\!\mu)t}{|n|}\\
\left(\frac{z}{z_0}\right)^{1\!-\!\mu}\!
&=1\!+\!i\frac{2(1\!-\!\mu)t}{|n|}
\end{split}
\label{eq:class}
\end{equation}
where we have reintroduced the complex coordinate vector $z$ 
($r^2=z\bar z$), and 
$z_0$ is its value for $t\!=\!0$.
Eqs.~(\ref{eq:class}) give the classical trajectory 
for the effective particle which agrees with previous
results \cite{Bellini:1996mb,Menotti:1999pn}.
One sees that in the conformal limit of small $1\!-\!\mu$, 
the angle $\theta$ grows 
linearly with time while 
the radius $r$ is fixed: we have a circular motion
of body 2 around body 1.

\section{Outlook}

The preceding analysis shows that in the two body case, the Hamiltonian
of both 2+1-Gravity and the high energy model are related to a basic
scale of the problem which, in the conformal limit, takes the form
$\sim\log|p|^2|z|^2$. Thus canonical quantization, with proper
ordering of operators, leads to analogous features,
even when scaling violations $O(1\!-\!\mu)$ are turned on.

A natural question is whether this analogy is kept in the
many-body case. In the conformal limit of the high-energy model,
some exact three-body solutions are known (the ``odderon'' 
\cite{Lipatov:1993qn,Janik:1998xj}).
Much less is known in the case of running coupling
(or scaling violations).
On the other hand, the general structure of the Hamiltonian
is known \cite{Cantini:2000dd} in 2+1-Gravity too, 
but becomes tractable
in the small speed limit only \cite{Bellini:1996mb}.
Furthermore, the conformal limit $\mu\!\rightarrow\!1$ is
ambiguous, in the sense that it depends on the mass parameters
$\mu_i$ also, and it is conceivable that the high-energy
model may correspond to some special mass configuration.
Therefore, a deeper analysis is needed in order to understand
whether the amusing correspondence found here survives in the
general case.

\section*{Acknowledgments}

We are grateful to Andrea Cappelli and Domenico Seminara
for interesting discussions.


\begin{thebibliography}{10}

\bibitem{Kuraev:1977fs}
E.A. Kuraev, L.N. Lipatov and V.S. Fadin,
\newblock Sov. Phys. JETP 45 (1977) 199.

\bibitem{Balitsky:1978ic}
I.I. Balitsky and L.N. Lipatov,
\newblock Sov. J. Nucl. Phys. 28 (1978) 822.

\bibitem{Maldacena:1998re}
J.M. Maldacena,
\newblock Adv. Theor. Math. Phys. 2 (1998) 231, hep-th/9711200.

\bibitem{Strominger:2001pn}
A. Strominger,
\newblock JHEP 10 (2001) 034, hep-th/0106113.

\bibitem{Deser:1984tn}
S. Deser, R. Jackiw and G. 't~Hooft,
\newblock Ann. Phys. 152 (1984) 220.

\bibitem{Deser:1984dr}
S. Deser and R. Jackiw,
\newblock Annals Phys. 153 (1984) 405.

\bibitem{Klemm:2002ir}
D. Klemm and L. Vanzo,
\newblock JHEP 04 (2002) 030, hep-th/0203268.

\bibitem{Bellini:1995rw}
A. Bellini, M. Ciafaloni and P. Valtancoli,
\newblock Nucl. Phys. B454 (1995) 449, hep-th/9507077.

\bibitem{Menotti:1999pn}
P. Menotti and D. Seminara,
\newblock Annals Phys. 279 (2000) 282, hep-th/9907111.

\bibitem{Deser:1988qn}
S. Deser and R. Jackiw,
\newblock Commun. Math. Phys. 118 (1988) 495.

\bibitem{'tHooft:1988yr}
G. 't~Hooft,
\newblock Commun. Math. Phys. 117 (1988) 685.

\bibitem{Carlip:1993zi}
S. Carlip,
\newblock Canadian Gen.Rel.1993:0215-234  (1993), gr-qc/9305020.

\bibitem{Witten:1988hc}
E. Witten,
\newblock Nucl. Phys. B311 (1988) 46.

\bibitem{Achucarro:1986vz}
A. Achucarro and P.K. Townsend,
\newblock Phys. Lett. B180 (1986) 89.

\bibitem{ADM}
R. Arnowitt, S. Deser and C. Misner,
\newblock edited by L. Witten and John Wiley and Sons  (1962).

\bibitem{Moncrief:1989dx}
V. Moncrief,
\newblock J. Math. Phys. 30 (1989) 2907.

\bibitem{Bellini:1996mb}
A. Bellini, M. Ciafaloni and P. Valtancoli,
\newblock Nucl. Phys. B462 (1996) 453, hep-th/9511207.

\bibitem{Welling:1996er}
M. Welling,
\newblock Class. Quant. Grav. 13 (1996) 653, hep-th/9510060.

\bibitem{Cantini:2000dd}
L. Cantini, P. Menotti and D. Seminara,
\newblock Class. Quant. Grav. 18 (2001) 2253, hep-th/0011070;
\newblock hep-th/0203103.

\bibitem{Lipatov:1990zb}
L.N. Lipatov,
\newblock Phys. Lett. B251 (1990) 284.

\bibitem{Lipatov:1993qn}
L.N. Lipatov,
\newblock Phys. Lett. B309 (1993) 394.

\bibitem{Faddeev:1995zg}
L.D. Faddeev and G.P. Korchemsky,
\newblock Phys. Lett. B342 (1995) 311, hep-th/9404173.

\bibitem{Fadin:1998py}
V.S. Fadin and L.N. Lipatov,
\newblock Phys. Lett. B429 (1998) 127, hep-ph/9802290.

\bibitem{Ciafaloni:1998gs}
M. Ciafaloni and G. Camici,
\newblock Phys. Lett. B430 (1998) 349, hep-ph/9803389.

\bibitem{Camici:1997ij}
G. Camici and M. Ciafaloni,
\newblock Phys. Lett. B412 (1997) 396, hep-ph/9707390.

\bibitem{Janik:1998xj}
R.A. Janik and J. Wosiek,
\newblock Phys. Rev. Lett. 82 (1999) 1092, hep-th/9802100.

\end{thebibliography}

\end{document}